\documentclass[twocolumn]{aa}

\usepackage{hyperref}
\usepackage{amsmath,amssymb,mathrsfs}

\usepackage{graphicx,natbib,verbatim}
\usepackage[varg]{txfonts}

\bibpunct{(}{)}{;}{a}{}{,}

\begin{document}

\titlerunning{Measuring X-ray anisotropy in solar flares}

\authorrunning{Casadei, Jeffrey \& Kontar}

\title{\textbf{Measuring X-ray anisotropy in solar flares. Prospective stereoscopic capabilities of STIX and MiSolFA}}

\author{Diego Casadei\inst{1}, Natasha L. S. Jeffrey\inst{2} \and Eduard P. Kontar\inst{2}}

\institute{University of Applied Sciences \& Arts Northwestern
  Switzerland (FHNW), Bahnhofstrasse 6, 5210 Windisch, Switzerland
  \and
  School of Physics \& Astronomy, University of Glasgow, Glasgow G12 8QQ, UK
} 

\offprints{{Diego Casadei \email{diego.casadei@fhnw.ch}}}

\date{Received ; Accepted}

\abstract{
  During a solar flare, a large percentage of the magnetic energy
  released goes into the kinetic energy of non-thermal particles, with
  X-ray observations providing a direct connection to keV
  flare-accelerated electrons.  However, the electron angular
  distribution, a prime diagnostic tool of the acceleration mechanism
  and transport, is poorly known.
} {
  During the next solar maximum, two upcoming space-borne X-ray
  missions, STIX on board \emph{Solar Orbiter} and MiSolFA, will
  perform stereoscopic X-ray observations of solar flares at two
  different locations: STIX at 0.28 AU (at perihelion) and up to
  inclinations of $\sim25^{\circ}$, and MiSolFA in a low-Earth orbit.
  The combined observations from these cross-calibrated detectors will
  allow us to infer the electron anisotropy of individual flares
  confidently for the first time.
} {
  We simulated both instrumental and physical effects for STIX and
  MiSolFA including thermal shielding, background and X-ray Compton
  backscattering (albedo effect) in the solar photosphere. We predict
  the expected number of observable flares available for stereoscopic
  measurements during the next solar maximum.  We also discuss the
  range of useful spacecraft observation angles for the challenging
  case of close-to-isotropic flare anisotropy.
} {
  The simulated results show that STIX and MiSolFA will be capable of
  detecting low levels of flare anisotropy, for M1-class or stronger
  flares, even with a relatively small spacecraft angular separation
  of 20--30$^{\circ}$. Both instruments will directly measure the
  flare X-ray anisotropy of about 40 M- and X-class solar flares
  during the next solar maximum.
} {
  Near-future stereoscopic observations with \emph{Solar Orbiter}/STIX
  and MiSolFA will help distinguishing between competing
  flare-acceleration mechanisms, and provide essential constraints
  regarding collisional and non-collisional transport processes
  occurring in the flaring atmosphere for individual solar flares.
}


\maketitle

\section{Introduction}\label{intro}

Solar flares are the most powerful explosive events in the solar
system, and large flares can release up to $10^{32}$ ergs of energy in
a few minutes \citep{2008LRSP....5....1B,2011SSRv..159..107H}.  A
large fraction of this energy, stored in coronal magnetic fields and
released by magnetic reconnection, goes into the acceleration of
particles.  However, the mechanisms transforming magnetic energy into
kinetic energy are still not clearly understood.  Flare-accelerated
electrons emit a continuous spectrum of bremsstrahlung X-rays that can
span a wide energy range up to gamma rays.  Hard X-rays (HXR,
$\gtrsim 20$ keV) are a direct link to flare-accelerated electrons and
a vital probe of the flare physical processes occurring at the Sun
\citep[e.g.][]{2003ApJ...595L.115B,2011SSRv..159..301K,2011SSRv..159..107H}.

Current X-ray observations are performed by the space-borne
\emph{Reuven Ramaty High Energy Solar Spectroscopic Imager}
\citep[RHESSI;][]{2002SoPh..210....3L}, using nine rotation modulation
collimators and bulk germanium (Ge) detectors to perform indirect
imaging and spectroscopy from 3 keV to 17 MeV with an angular
resolution of few arcseconds \citep{2002SoPh..210...61H}.  The
brightest X-ray sites are often found at the footpoints of newly
reconnected magnetic loops that link coronal acceleration regions with
the much denser chromosphere.  Here, the bulk of the accelerated
electrons interact, losing energy via electron-electron Coulomb
collisions and emitting bremsstrahlung X-rays by interacting with the
ambient ions.
Solar flare X-ray observations typically show a superposition of two
distributions.  The first is a thermal component emanating from the
corona, with flare temperatures of a few tens of million degrees,
dominating up to 10-20 keV.\ The second is a non-thermal power law
extending to higher energies with a spectral index in the range from
$\sim$2 to 5 for HXR footpoints sources and, when detected, from
$\sim$3 to 8 for the coronal non-thermal emission
\citep{2003ApJ...595L.107E,2005SoPh..232...63K,2008ApJ...673.1181K,2013A&A...551A.135S,2013ApJ...777...33C}.

Although the flare X-ray energy spectrum is well observed by RHESSI,
the angular distribution is poorly constrained.  The X-ray spectrum is
dependent on the angular distribution of the parent electrons
\citep[e.g.][]{2004ApJ...613.1233M,2011SSRv..159..301K} and
uncertainty regarding the electron pitch-angle distribution can also
lead to changes in inferred plasma parameters, e.g.\ an overestimation
of coronal density from X-ray imaging
\citep{2014ApJ...787...86J}. Thus, knowing the directivity of both the
injected and radiating electron distributions is essential for
understanding the type of acceleration mechanism(s) and the transport
and interactions of solar flare electrons.  Often, in an
oversimplified collisional thick-target model, the injected and
emitting electrons are assumed to be beamed along the guiding field
lines \cite[e.g.][]{1971SoPh...18..489B}.  However acceleration (for
example by a second-order Fermi process) might produce an isotropic
distribution of accelerated electrons
\citep[e.g.][]{1994ApJS...90..623M,1996ApJ...461..445M,2012SSRv..173..535P}.
Furthermore, electron transport through the surrounding solar plasma
ultimately broadens the electron distribution, increasing the isotropy
by collisional or non-collisional pitch-angle scattering
\citep{2014ApJ...780..176K}.  Therefore, even if the injected
distribution is strongly beamed, the angular distribution of radiating
electrons is expected to isotropise as they are transported from the
corona to the chromosphere.

So far, HXR directivity has been studied with the following techniques
\citep[e.g.][as a review]{2011SSRv..159..301K}:
\begin{enumerate}
\item\label{method-1} Statistical flare studies of centre-to-limb
  variations in flux or spectral index
  \citep[e.g.][]{1969SoPh....7..260O,2007A&A...466..705K}

\item\label{method-2} Albedo mirror analysis of strong solar flares
  \citep{2006ApJ...653L.149K, 2013SoPh..284..405D}

\item\label{method-3} Linear X-ray polarization measurements from a
  single flare with one satellite
  \citep[e.g.][]{1970SoPh...14..204T,2004AdSpR..34..462M,2006SoPh..239..149S}

\item\label{method-4} Simultaneous observations of a single flare with
  two satellites at different viewing angles
  \citep[e.g.][]{1981Ap&SS..75..163K,1986ASSL..123...73H,1988ApJ...326.1017K,1990ApJ...359..524M,1998ApJ...500.1003K}.
\end{enumerate}

Method~\ref{method-2} suggests that the HXR emitting electron
distribution is close to isotropic, and not beamed as in a simple
standard flare model, at least for the few events published. This
method uses the X-ray albedo effect
\citep[e.g.][]{1972ApJ...171..377T,1973SoPh...29..143S,1978ApJ...219..705B},
where sunwards emitted X-rays are Compton backscattered in the
photosphere into the observer direction.  The directivity is then
determined by separating the directly emitted and reflected components
of HXR flux that contribute to the observed X-ray spectrum.
\citet{2006ApJ...653L.149K} studied two flares and their
analysis showed that both flares were close to isotropic. A follow-up
study of eight events by \citet{2013SoPh..284..405D} again found a
lack of electron anisotropy below 150 keV. 

Recently \citet{2007A&A...466..705K} studied 398 flares using
method~\ref{method-1}, accounting for the albedo component.  Although
they found changes in spectral index that were consistent with the
presence of an albedo component, the statistical study gave no clear
conclusion regarding average flare directivity.

Method~\ref{method-3} uses the direct link between X-ray linear
polarization and electron anisotropy.  Electron directivity and X-ray
polarization have been extensively modelled (with and without albedo)
with different scenarios
\citep[e.g.][]{1983ApJ...269..715L,1978ApJ...219..705B,2008ApJ...674..570E,2011A&A...536A..93J}.
Nevertheless, observations with past instruments and non-dedicated
polarimeters, such as RHESSI, have proved inconclusive, owing to
instrumental issues (small effective area etc.) inducing large
uncertainties associated with the measurements.
The recently launched POLAR \citep{2015JAMP..3..272H}, a wide
field-of-view X-ray polarimeter installed on the Chinese space
station \emph{Tiangong-2} in September 2016, should allow for
confident detection of X-ray polarization during large flares at
suitable heliocentric angles away from the solar disk
centre. Importantly, POLAR has an effective area of 200 cm$^{2}$,
i.e.\ two orders of magnitude larger than the polarimeter on board
RHESSI, and a low minimum detectable polarization of 5\%, that should
remove some of the previous issues (e.g. high background levels and large
uncertainties). However, POLAR will be operational during a period of
decreasing solar activity.

Unambiguous measurements of solar flare electron anisotropy can be
obtained through X-ray directivity measurements made by
cross-calibrated detectors looking at the same source from two
separate points of view (method~\ref{method-4}).  Such previous stereoscopic
studies \citep[e.g.][]{1998ApJ...500.1003K} found no clear evidence
for directivity at large X-ray energies.  However, past direct
measurements by multiple spacecrafts suffered greatly from calibration
issues, owing to the use of different types of detectors, therefore the
results were questionable at best.  Thus, it is fundamental that the
two instruments have a well-known energy cross-calibration.

A concrete possibility of obtaining simultaneous stereoscopic
observations will be provided by two instruments, which will operate
at the next solar maximum.  The ``Spectrometer Telescope for Imaging
X-rays'' (STIX) \citep{2013NIMPA.732..295K}, is an instrument to be
flown on board the ESA/NASA \emph{Solar Orbiter} mission
\citep{2013SoPh..285...25M} within the ESA Cosmic Vision programme.
\emph{Solar Orbiter} will be launched in October 2018 and it will
start its science programme after a three year cruise phase.
The ``Micro Solar-Flare Apparatus'' (MiSolFA) \citep{casadei2014} is a
compact X-ray detector being developed in Switzerland, in
collaboration with the French STIX team and the Italian Space Agency,
to be operated in low-Earth polar orbit at the next solar maximum.
Thus, during the next solar maximum period simultaneous observations
will be jointly performed by STIX and MiSolFA, the first orbiting
around the Sun and the second around the Earth.  Importantly, both
instruments will adopt the same type of photon detectors, overcoming
the calibration issues of past instruments.

The purpose of this paper is to discuss the prospective stereoscopic
observations with STIX and MiSolFA, highlighting the capabilities of
both instruments working in tandem, and estimating the number of
observable flares, which are suitable for measuring the anisotropy of
the X-ray emission.
The two instruments are illustrated in Section
\ref{dual_obs}. Simulated solar flares with small differences in
electron anisotropy are studied, and we consider both instrumental and
physical effects such as X-ray albedo. In Section \ref{sims} it is
shown that, even with a spacecraft angular separation as small as
20--30$^{\circ}$, a mildly anisotropic distribution will produce
detectable differences in the observed X-ray spectra, for a bright
enough solar flare. The expected number of flare observations is
finally estimated in Section \ref{exp_flares}, where it is found that
several flares per year will be suitable for an energy-dependent
directivity measurement with STIX and MiSolFA.  In Section
\ref{summary}, all the main results are summarized.

\section{Dual observations with STIX and MiSolFA}\label{dual_obs}

The STIX instrument will provide X-ray imaging spectroscopy from 4 to
150~keV with 32 Caliste-SO units equipped with CdTe crystals
\citep{2012NIMPA.695..288M} and energy resolution better than 1~keV
(FWHM) from 14 to 60~keV.  The imaging is performed with an indirect
technique based on the Moir\'{e} effect, achieving an angular
resolution of about 7 arcseconds.  At perihelion, STIX will observe
flares from a distance three times closer to the Sun than instruments
orbiting around the Earth, hence achieving an effective spatial
resolution similar to RHESSI (which has $\sim2$ arcsecond resolution).
In order to complement STIX observations with minimal differences in
the energy response, MiSolFA will adopt the same photon detectors,
i.~e.\ Caliste units equipped with CdTe crystals with 1~mm thickness.
The orbital inclination of STIX will increase over time up to
$25^{\circ}$ or more (depending on the mission duration).  Hence, the
two instruments will be able to observe the same flare
stereoscopically from two different points of view.

The STIX and MiSolFA instruments both need to exploit indirect imaging
techniques because they cannot accommodate grazing-incidence focussing
optics, which require large focal distances of several metres.  For
example, astronomical direct X-ray imagers such as the \emph{Nuclear
  Spectroscopic Telescope Array} (NuSTAR) have a 10~m focal length
\citep{2013ApJ...770..103H}.  Like RHESSI, STIX and MiSolFA rely on
the Moir\'{e} effect, which is produced by a pair of parallel grids
placed in front of each photon detector.
The STIX instrument has 30 pairs of tungsten grids providing Fourier
components with 9 different directions (at 20$^{\circ}$ steps) and 10
angular scales (from 7 to 950 arcseconds, with constant-ratio steps of
$\sqrt{2}$).  Hence, STIX will be able to precisely locate the flare
on the Sun and study the morphology of the X-ray emitting sources.  On
the other hand, the main purpose for the imaging system of MiSolFA is
to separate the flare HXR footpoints, relying on other observations to
locate the source on the Sun. For this purpose, it is sufficient to
sample a relatively narrow angular range (from 10 to 60
arcseconds). MiSolFA will cover this range with 12 subcollimators
sampling two orthogonal directions with frequencies following a
Fourier series. Therefore, although only in a very limited angular
range, MiSolFA will have a better point spread function than STIX,
whereas the latter will be able to cover a much wider angular range
and will be able to locate the source with very high precision (of
order of 1 arcsecond).

In the rest of this paper, we focus on X-ray spectroscopy alone, hence
the imaging performance of the two instruments is ignored, apart from
taking into account the reduction of effective area (because of the
grids, only about 25\% of the photons reach the detectors).
The effective area of MiSolFA detectors is 40\% of the STIX area.  In
addition, the time average over the high-eccentricity orbit of STIX
gives it almost a factor 3 increase in intensity, owing to the closer
distance from the Sun.  This gives a ratio of 7.5 between STIX and
MiSolFA acceptances.
We also account for additional details in the following simulations.
Arriving photons ``see'' the thermal shield of each instrument, which
is designed to stop radiation up to the X-ray range.  The STIX
instrument has two beryllium (Be) windows with total thickness of
3~mm, whereas a 0.3~mm thick Be layer has been considered for MiSolFA.

\begin{figure}[t!]
   \centering
   \includegraphics[width=0.95\linewidth]{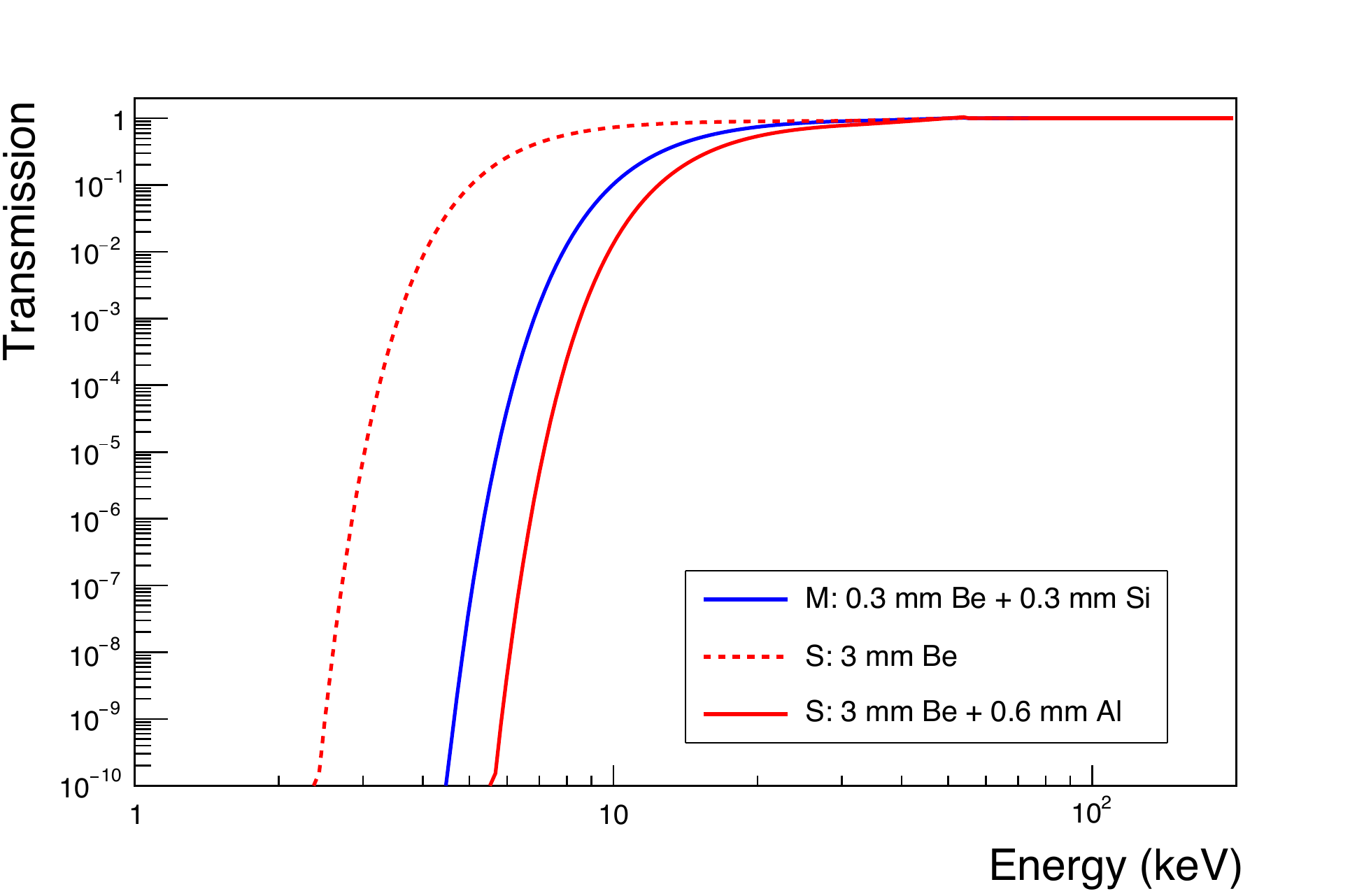}
   \caption{Transmission of the passive material in front of STIX (S)
     and MiSolFA (M) X-ray detectors.  STIX employs a movable Al
     attenuator with 0.6~mm thickness and two Be windows with total
     thickness of 3~mm, while MiSolFA has a 0.3~mm Si layer and
     a Be 0.3~mm layer.}
   \label{fig-transmission}
\end{figure}

In order to have sufficient counts up to about hundred keV, flares of
class M or above are considered.  This allows for the precise
determination of the thermal and non-thermal contributions to the
total flux.  To avoid large dead times owing to the high flux of
low-energy photons, STIX employs a movable aluminium attenuator with
0.6~mm thickness.  On the other hand MiSolFA has no movable part, as
this would compromise the pointing stability of such a light
satellite.  Nevertheless, the low-energy flux is also attenuated by
MiSolFA, which plans to adopt golden grids fabricated onto a silicon
or carbon substrate.  Here a Si layer is considered (a conservative
assumption), which absorbs most photons below 8~keV.  Hence the
MiSolFA transmission is not very different from what STIX achieves
with the attenuator in front of the photon detectors (see
figure~\ref{fig-transmission}).

Furthermore, although STIX and MiSolFA will adopt the same photon
detectors, their performance cannot be expected to be identical, since
\emph{Solar Orbiter} will only start collecting science data after a
cruise of three years.  During this initial phase, the CdTe crystals
will experience some ageing owing to incident radiation.
The radioactive sources installed on these detectors will provide
continuous calibration data, which is fundamental to ensure a good
cross-calibration.  Here it is conservatively estimated that the STIX
resolution will worsen by a factor of two \citep[a larger effect than
what found by][]{Eisen2002176,Zanarini2004315}, whilst no ageing is
considered for MiSolFA.

\begin{figure}[t!]
   \centering
   \includegraphics[width=0.9\linewidth]{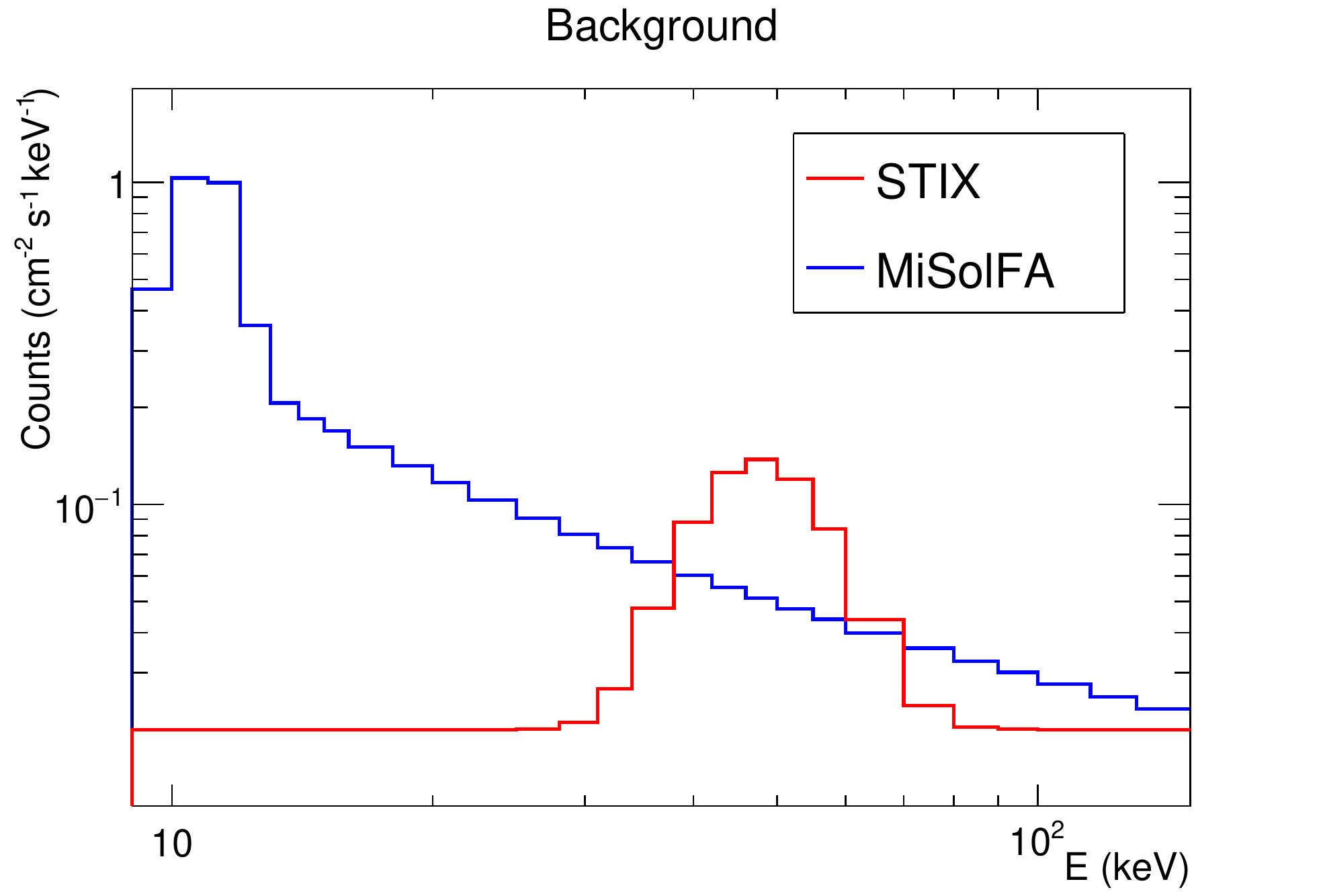}
   \caption{Estimated background counts for STIX and MiSolFA. The
     actual background counts for STIX and MiSolFA will be measured
     once both instruments are operating.}
   \label{fig-backgrounds}
\end{figure}

Finally, the background counts will have different distributions for
the two detectors.  One example is shown in
figure~\ref{fig-backgrounds}, where the MiSolFA background
distribution is taken to be similar to the RHESSI background at the
beginning of the mission, and for STIX a flat component is
superimposed with a bump that mimics the background X-ray radiation
\citep{1980ApJ...235....4M}. The actual background distributions will
be measured once the two instruments start operating.  Here what
matters is that they are expected to have different shapes, affecting
the measured photon spectra in different ways.

Flare spectra are rapidly falling energy distributions, while the
background counts have a more uniform distribution. Therefore, at high
energies (e.~g.\ well above 100~keV) the measurement will be
background dominated.  On the other hand, the low-energy part is
dominated by the thermal emission, which is not expected to show
significant anisotropy, although a thermal source can show low levels
(few percent) of directivity and polarization
\citep[e.g. see][]{1980ApJ...237.1015E}.  Hence the useful energy
range for directivity studies extends from the end of the thermal
region ($\sim20$~keV), up to the energy bins in which the background
counts are of the same order of magnitude as the actual flare photon
rate.  For example, for M-class flares this energy region roughly goes
from 20 to 100 keV.

\begin{figure}
   \centering
   \includegraphics[width=0.49\textwidth]{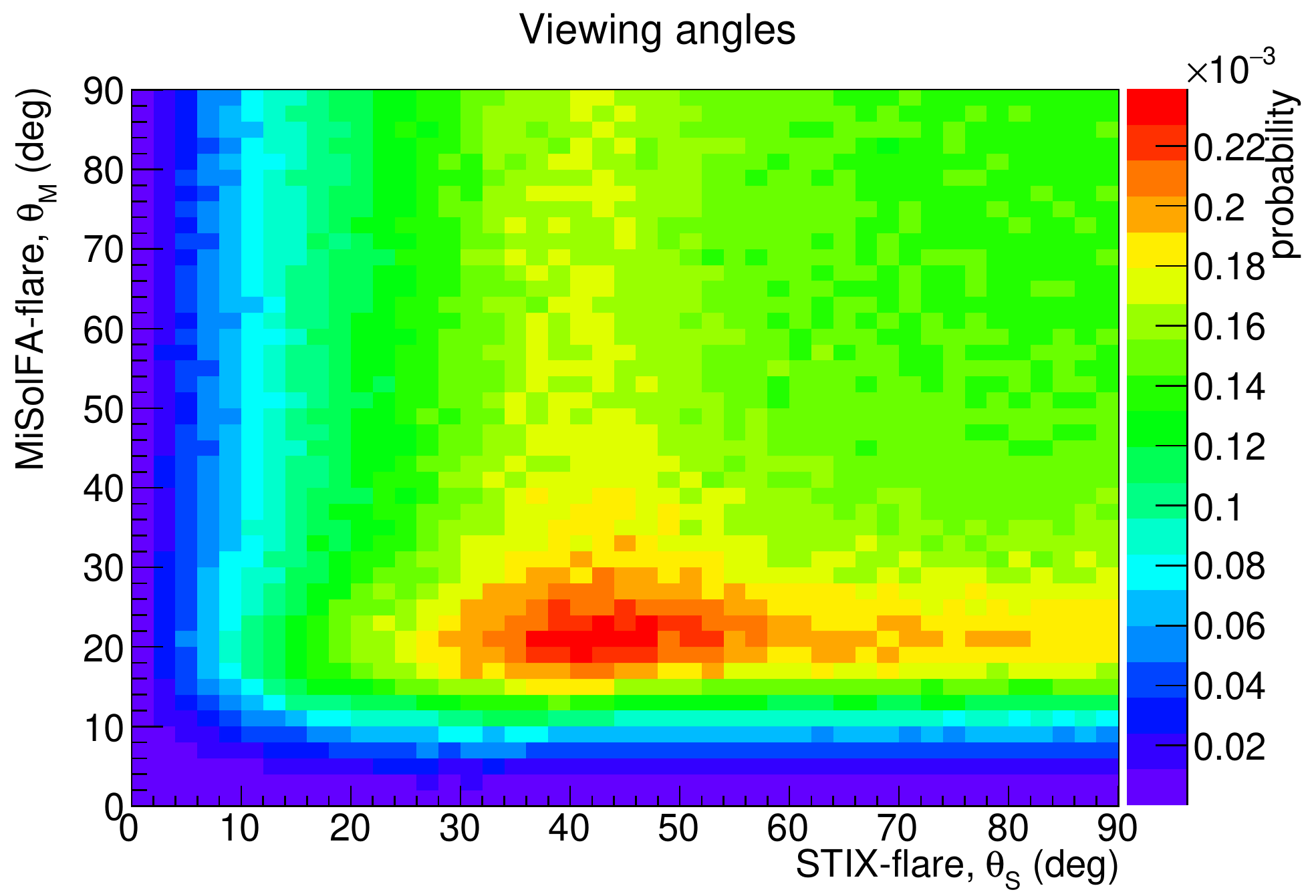}
   \caption{%
     Viewing angles of STIX ($\theta_\text{S}$) and MiSolFA
     ($\theta_\text{M}$).  The most probable ranges of viewing angles
     are $\theta_\text{S}\sim30$--$50^{\circ}$ and
     $\theta_\text{M}\sim20$--$30^{\circ}$.}
   \label{fig-angles}
\end{figure}

In order to estimate the flare viewing angles of STIX and MiSolFA, the
flare position on the Sun was simulated according to the measured
distribution seen by RHESSI over the past
decade.\footnote{\url{http://hesperia.gsfc.nasa.gov/hessidata/dbase/hessi_flare_list.txt}}
Similar to the sunspot distribution, flares are uniform in solar
longitude, which implies a higher density of flares approaching the
limb when looking from the Earth, and bimodal in latitude, with peaks
at about $\pm13.5^{\circ}$ and a root mean square of $6.3^{\circ}$.
With MiSolFA in a low Earth orbit and STIX taken
approximately\footnote{The orbit eccentricity is neglected here.}
uniform in latitude (within $\pm0.5$~rad) and in longitude (within
$\pm\pi$~rad), there is about 50\% probability (ignoring
beyond-the-limb flares\footnote{10--11\% of all flares observed by
  each instrument fall beyond the limb by no more than $10^{\circ}$,
  which leaves their coronal source visible. Beyond this angle a flare
  is fully occulted.}) that a flare is visible by one instrument,
hence the integral of the distribution in figure~\ref{fig-angles} is
0.25, which is the fraction of flares visible by both instruments.
There is a low probability for small viewing angles with a peak at
$\theta_\text{S}\sim30$--$50^{\circ}$ for STIX and
$\theta_\text{M}\sim20$--$30^{\circ}$ for MiSolFA; this is about 40\%
higher than a broad plateau reaching $90^{\circ}$, which corresponds
to limb flares.  Accounting for a 20\% live time for STIX,
corresponding to the fraction of time it will spend in science mode
during the main phase of \emph{Solar Orbiter}\footnote{This
  conservatively assumes no attempt to optimize the overlapping time
  with MiSolFA.} and for a fraction of flares visible by both
instruments equal to 50\%, one expects MiSolFA to observe the same
event as STIX for 10\% of all flares.  The MiSolFA instrument has a
two year nominal science mission duration, and this amounts to 73 days
of net observing time.  Assuming 20\% dead time, caused by some issue
for at least one of the instruments, one ends up with two months of
total simultaneous observing time.

\section{Simulation of flare measurement}\label{sims}

Starting from different electron distributions,
\citet{2011A&A...536A..93J} computed the X-ray bremsstrahlung emission
as a function of the electron energy and estimated the total X-ray
flux along all directions, for different degrees of electron
anisotropy.  Importantly, this work also included the X-ray albedo
component, where photons emitted towards the Sun are Compton
backscattered in the photosphere towards the observer with changed
photon properties.  The albedo component produces a bump in the energy
spectrum between $\sim$10--100 keV and this bump must be included
since it changes all the measured electron and photon properties,
including anisotropy.
In our simulation, we only modelled the non-thermal electron
distribution.  The resulting X-ray bremsstrahlung (including the
albedo component) is calculated for a HXR footpoint source located at
a chromospheric height of 1~Mm above the photosphere \citep[for more
information see][]{2010A&A...513L...2K,2011A&A...536A..93J}. The
energy spectrum of the emitting electrons is a single power law with
spectral index of 2, hence the injected electron distribution has a
spectral index of $\delta\sim4$ and the emitting photon distribution
has a spectral index of $\gamma\sim3$.  Here we make a comparison
between an isotropic and a mildly anisotropic case with a Gaussian
pitch-angle distribution with approximately 0.4~rad standard
deviation.

\begin{figure}[t]
  \centering
  \includegraphics[width=1.0\linewidth]{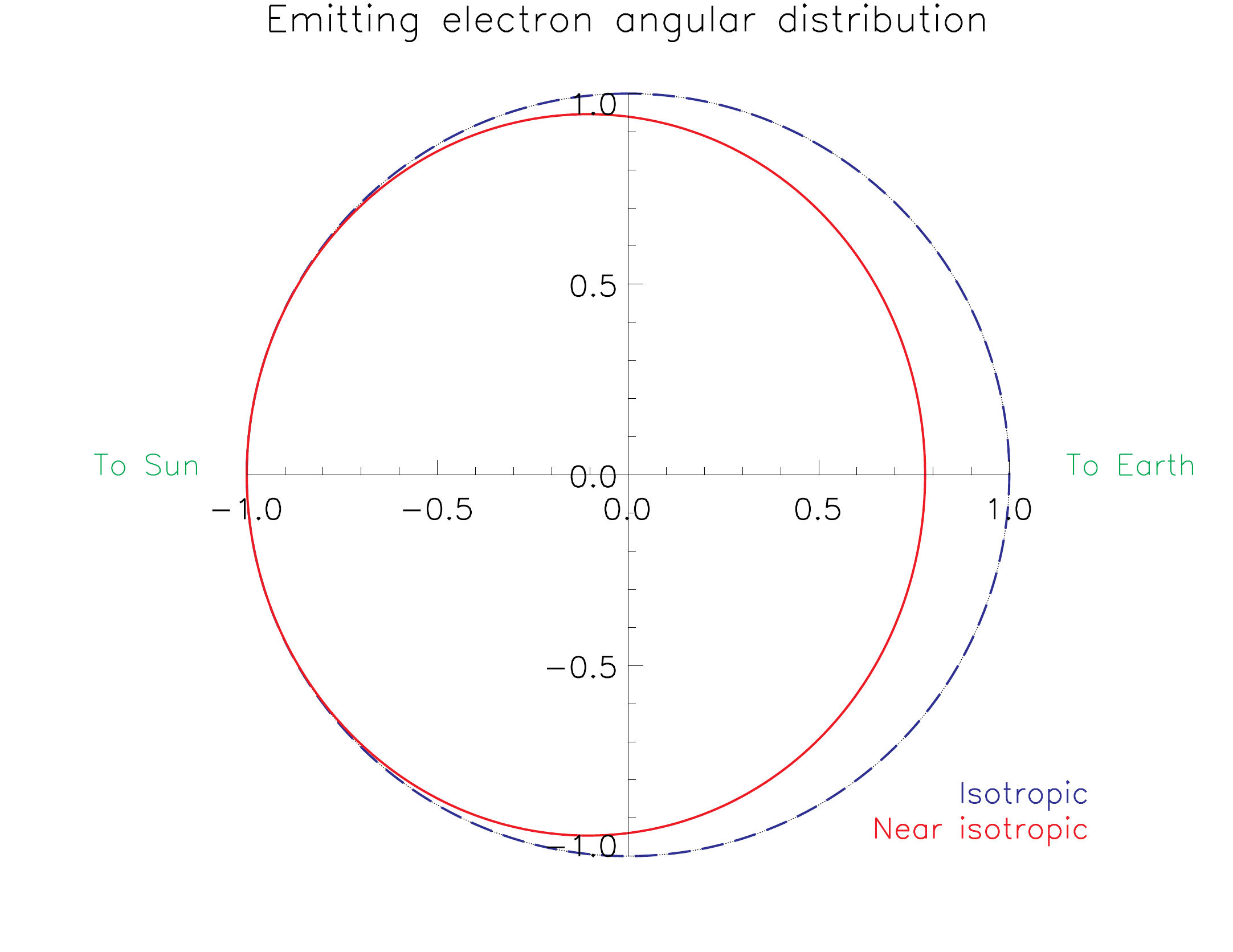}
  \caption{Polar plot of the simulated electron angular distributions
    for completely isotropic (blue) and mildly anisotropic (red). 
    In the simple model considered here, the energy dependence of the
    electron anisotropy is neglected.} 
  \label{fig-iso-near-iso}
\end{figure}

In the simulation, we use a simple model where the angular
distribution of the electrons does not change with energy, since more
complicated electron distributions are not required to show the
prospective stereoscopic capabilities of both instruments.  Here we
focus on the physically relevant observable differences in the
detected X-ray flux, which arise from different lines of sight,
albedo, and instrumental effects.  The latter are independent of the
details of the parent electron distribution, while both direct and
reflected emissions are expected to be functions of the electron
energy.  However, while the energy dependence of the albedo component
is well understood \citep{2011A&A...536A..93J}, the ``true'' electron
distribution is not known and likely changes in different flares.
Hence there was no attempt to model the full complexity of real
flares, as this is not necessary to assess the capability of the two
instruments to perform joint measurements from which the anisotropy
can be inferred, whatever the underlining electron model is. 
The angular distributions of the X-ray emitting electrons
used in our analysis are shown in Figure~\ref{fig-iso-near-iso}. The
non-isotropic distribution is only slightly sunwards beamed compared
to the isotropic case.

\begin{figure}[t]
  \centering
  \includegraphics[width=1.0\linewidth]{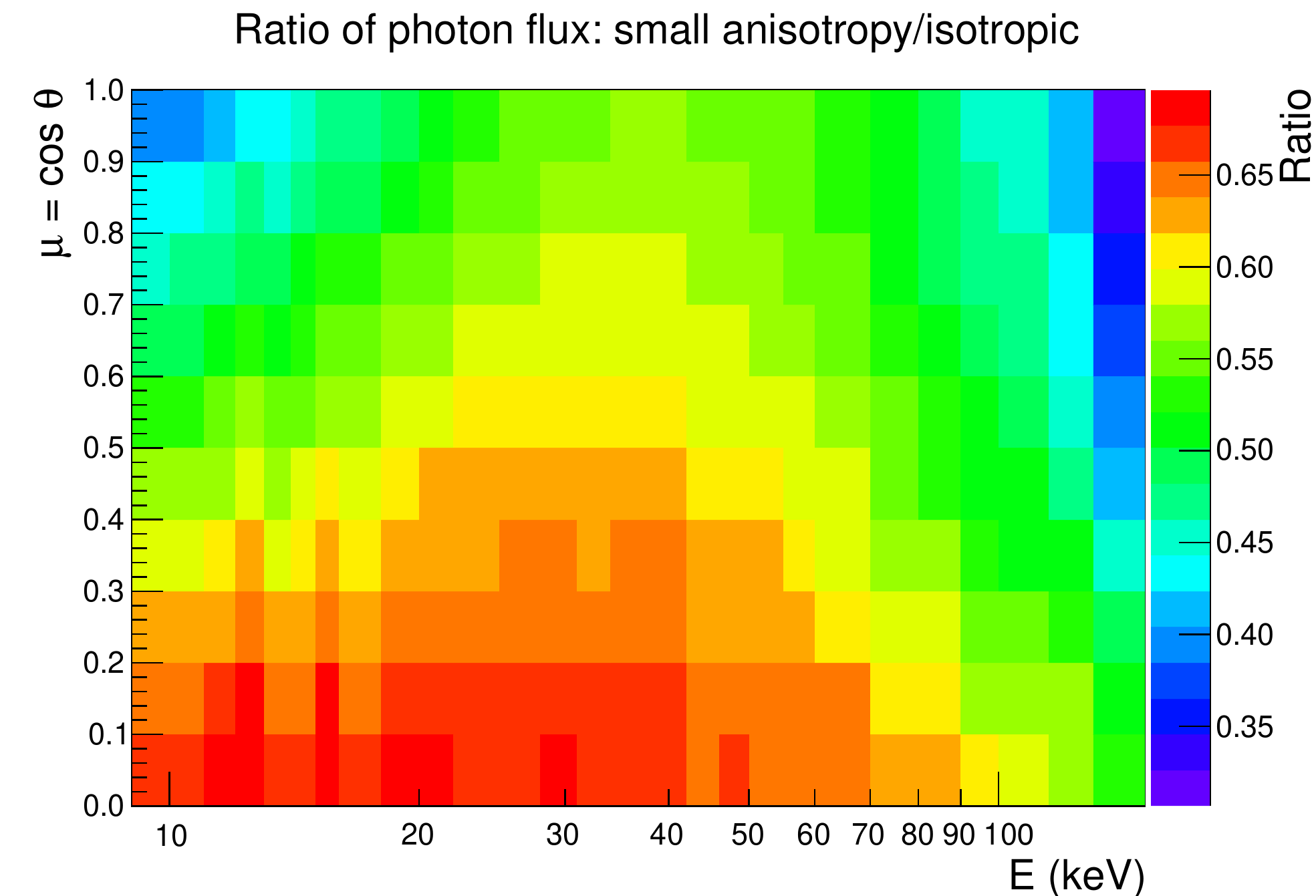}
  \caption{Binwise ratio between photons emitted by mildly anisotropic
    and fully isotropic electron distributions.  The isotropic thermal
    component is neglected in this plot.}
  \label{fig-true-ratio}
\end{figure}

\begin{figure}[t]
  \centering
  \includegraphics[width=1.0\linewidth]{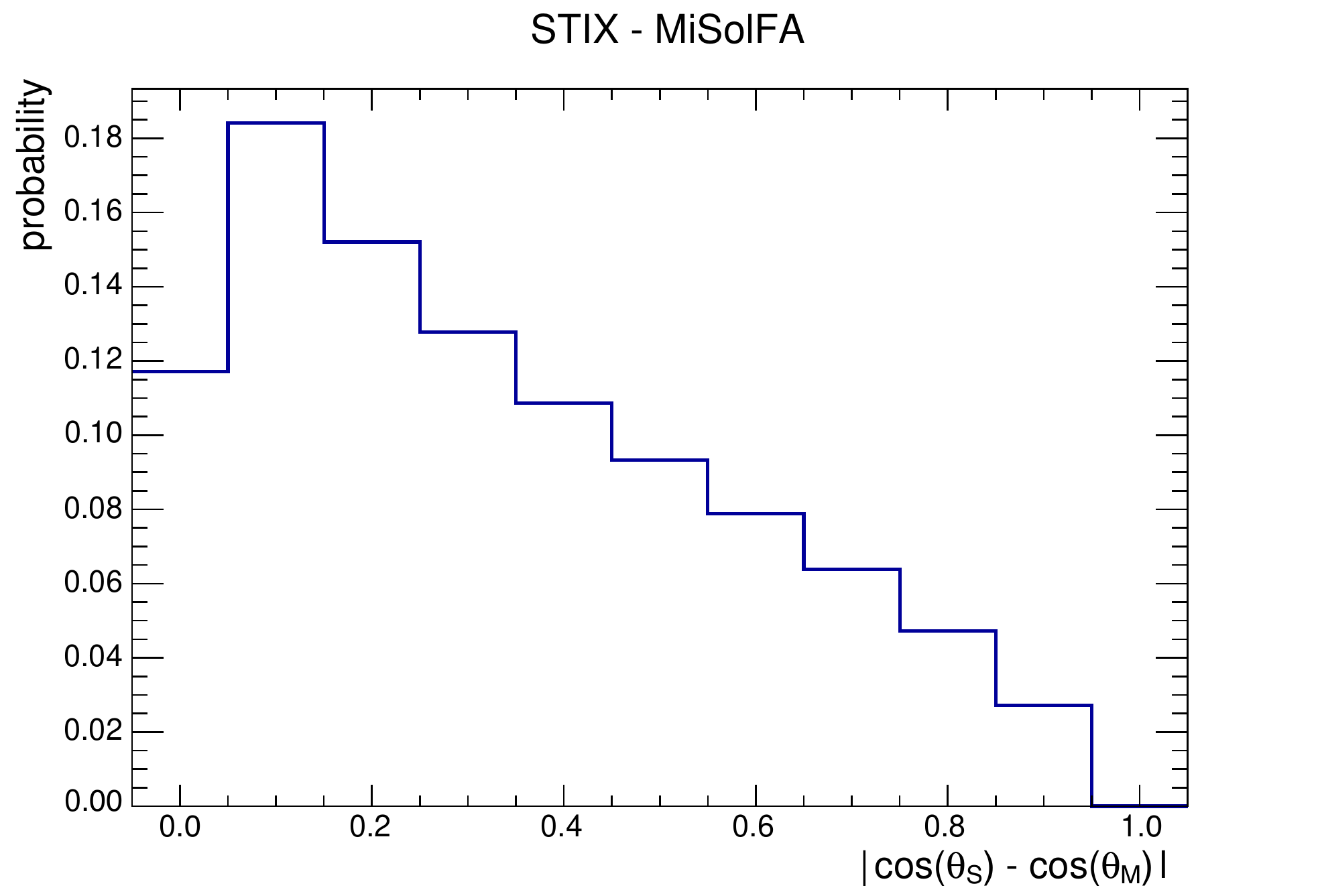}
  \caption{Difference in viewing angles.} 
  \label{fig-angle-diff}
\end{figure}

The two corresponding photon spectra have different energy dependence
along different directions.  Figure~\ref{fig-true-ratio} shows their
binwise ratio as a function of the energy and $\mu \equiv \cos\theta$,
where $\theta=\theta_\text{S}$ or $\theta_\text{M}$ is the viewing
angle with respect to the local flare vertical direction (equal to the
local heliocentric angle on the solar disk).  This means that, unless
STIX and MiSolFA are symmetrically located with respect to the flare
direction (which is very unlikely, see figure~\ref{fig-angles}), they
will measure different energy spectra.
Figure~\ref{fig-angle-diff} shows the difference
$|\cos\theta_\text{S}-\cos\theta_\text{M}|$ between STIX and
MiSolFA: most flares will be observed with angular separations large
enough to measure sizeable differences in the photon spectra.
Here we take $\cos\theta_\text{S}=$ 0.7--0.8 for STIX and
$\cos\theta_\text{M}=$ 0.9--1.0 for MiSolFA, to make a concrete
example. We used this conservative example since a larger difference
in the viewing angles will generally produce a greater difference in
the resulting energy spectra observed by both instruments.

\begin{figure*}[t!]
   \centering
   \includegraphics[width=0.5\textwidth]{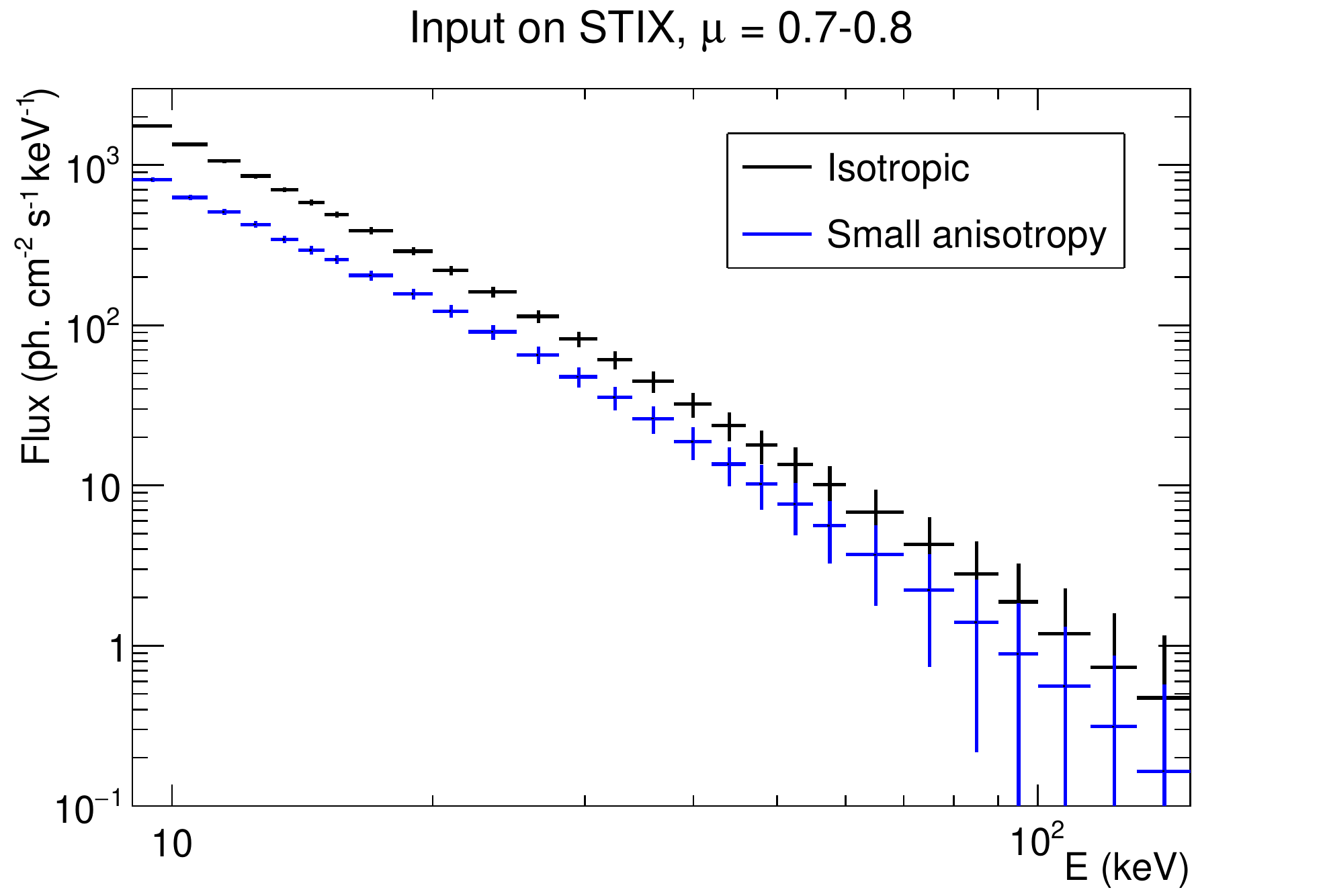}%
   \includegraphics[width=0.5\textwidth]{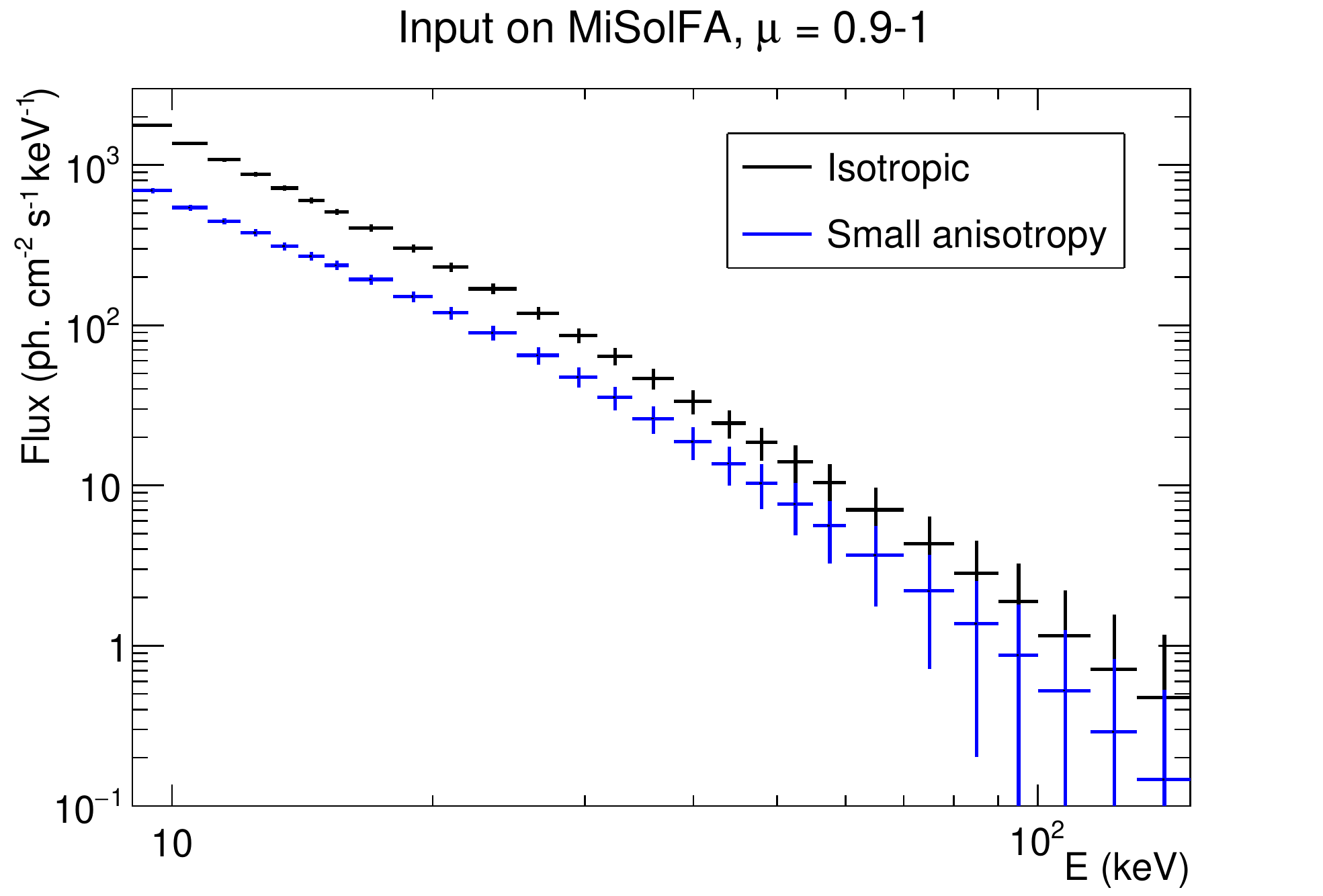}
   \caption{Simulated input spectra with $\cos\theta_\text{S}=$
     0.7--0.8 for STIX and $\cos\theta_\text{M}=$ 0.9--1.0 for
     MiSolFA.}
   \label{fig-true-flux}
   \vspace{2ex}

   \centering
   \includegraphics[width=0.5\textwidth]{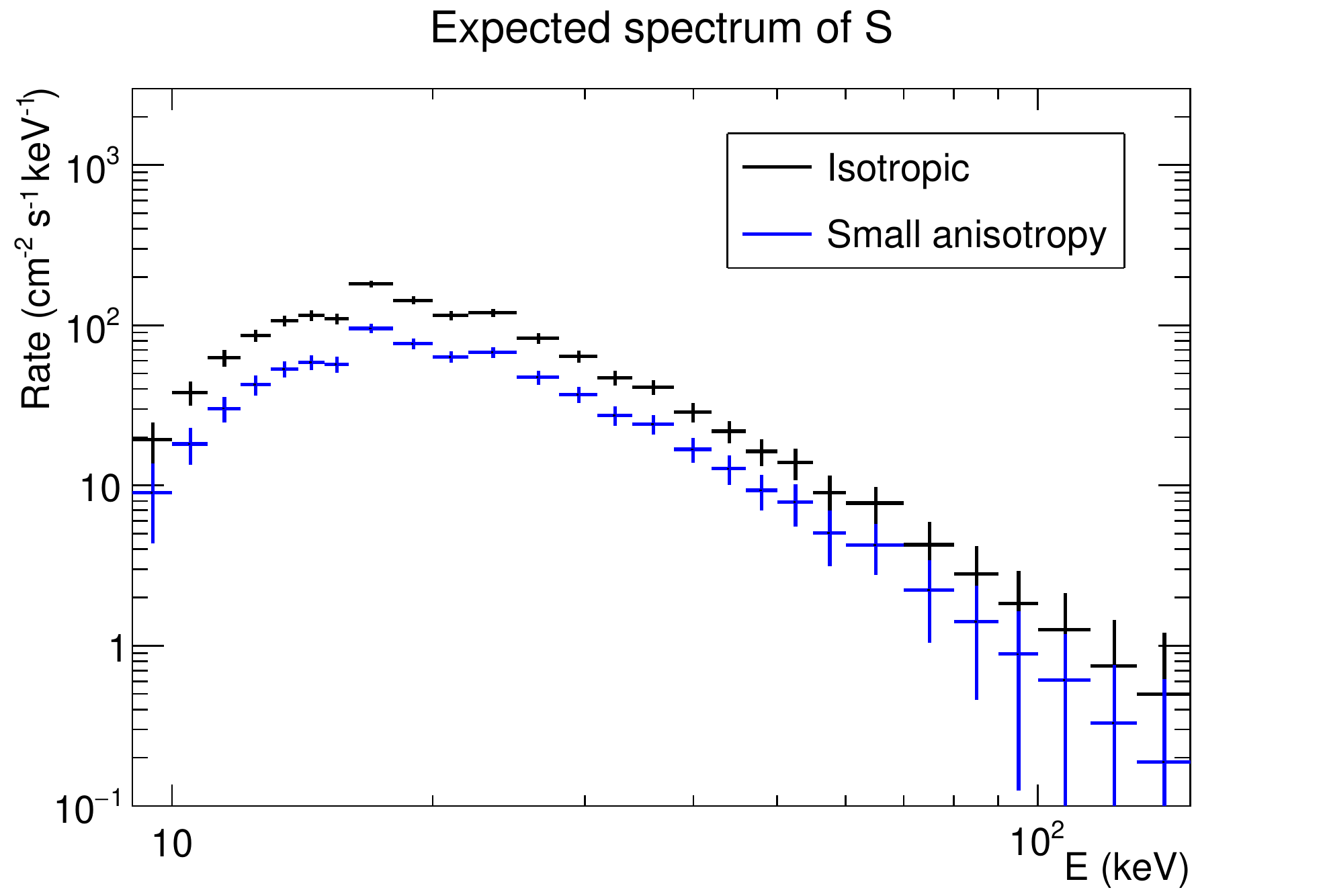}%
   \includegraphics[width=0.5\textwidth]{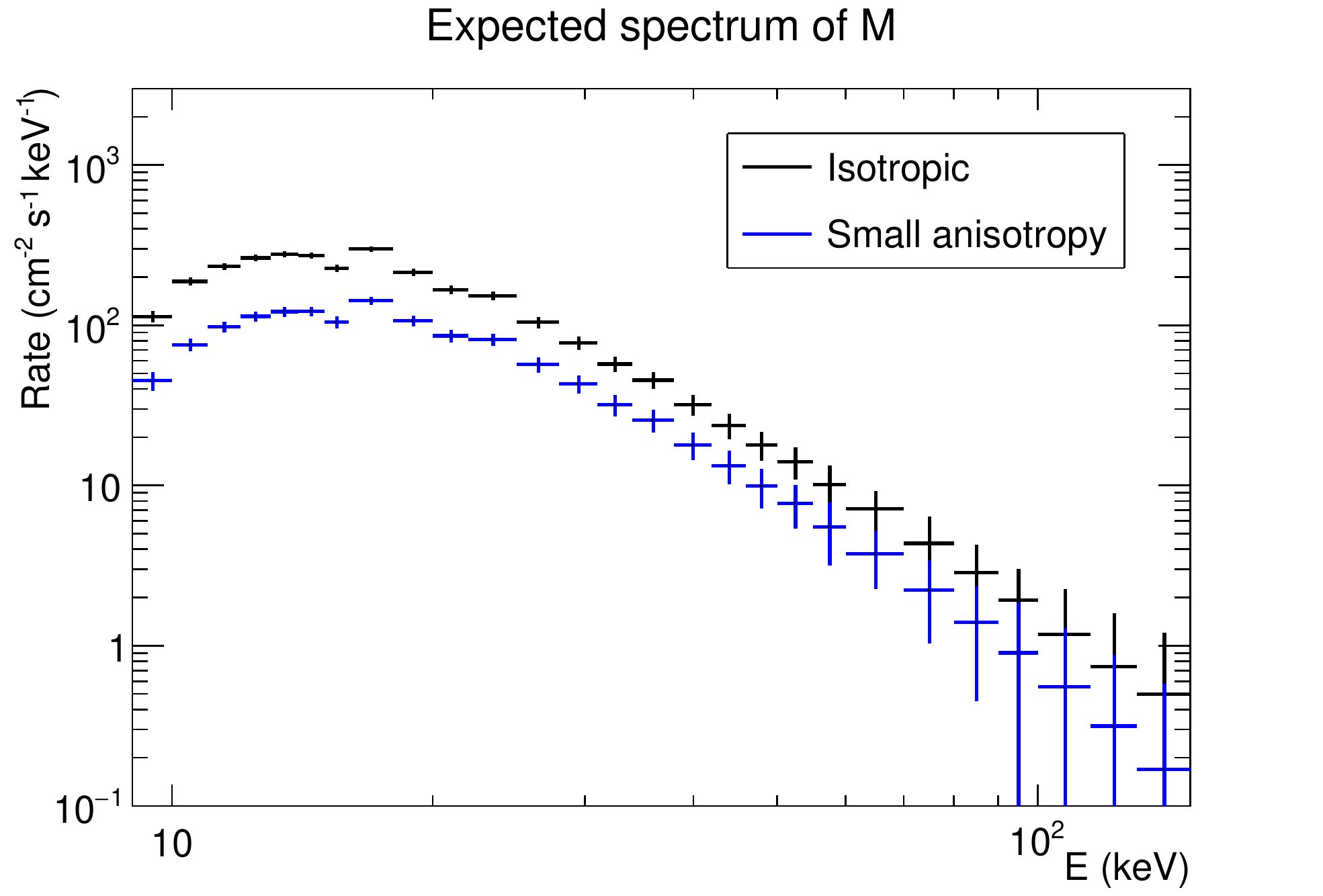}
   \caption{Expected energy spectra with $\cos\theta_\text{S}=$
     0.7--0.8 for STIX and $\cos\theta_\text{M}=$ 0.9--1.0 for
     MiSolFA.}
   \label{fig-expected-flux}
   \vspace{2ex}

   \centering
   \includegraphics[width=0.5\textwidth]{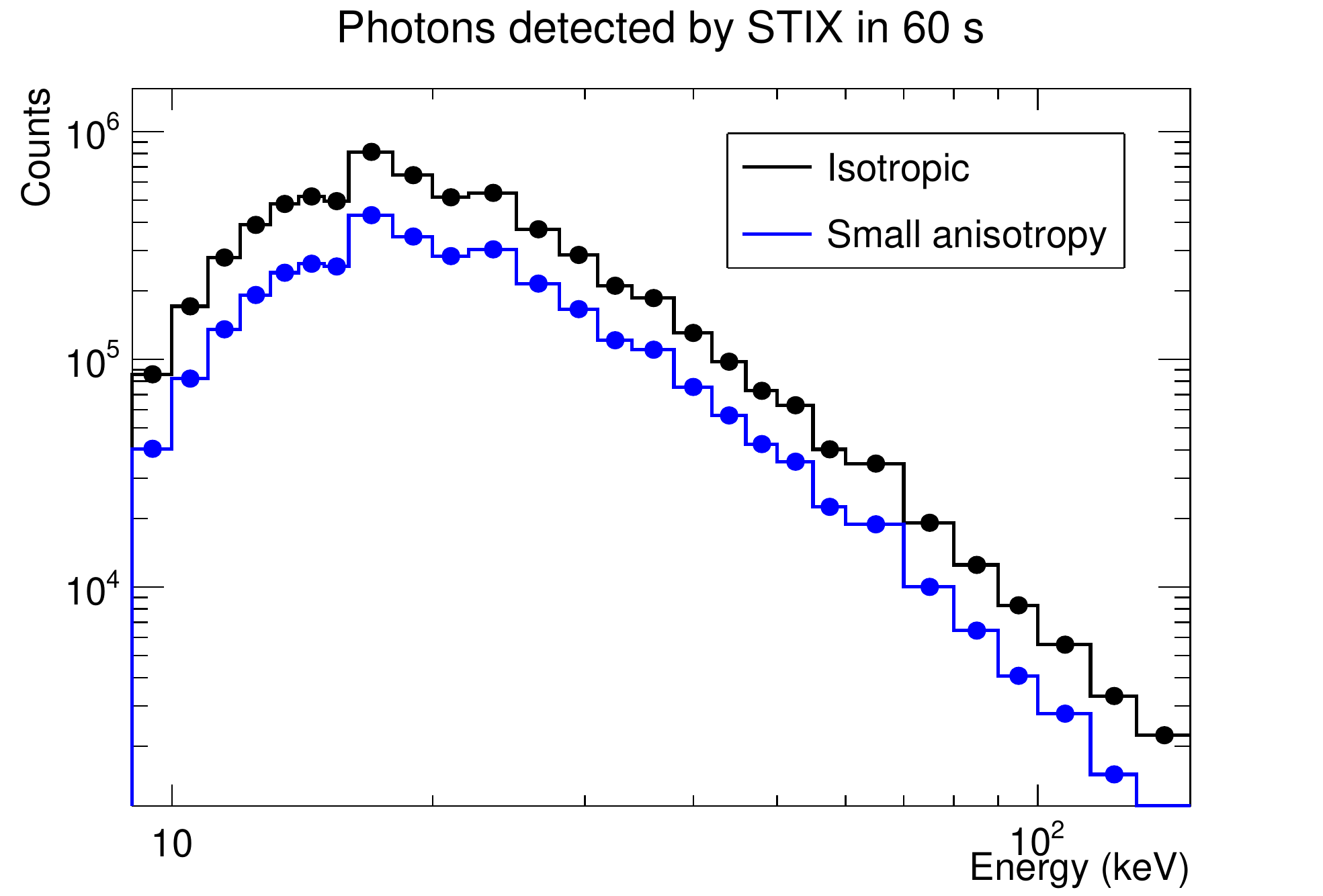}%
   \includegraphics[width=0.5\textwidth]{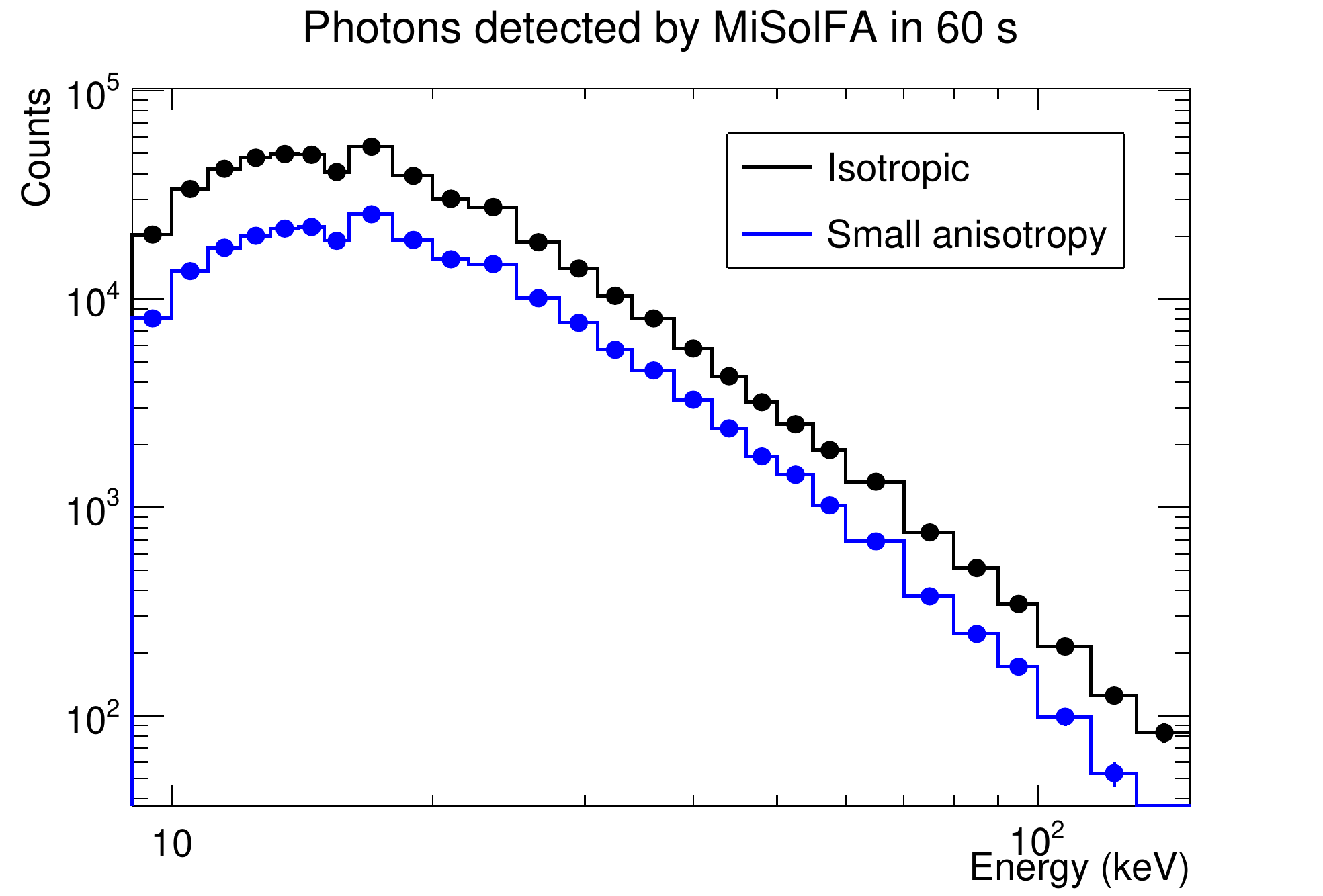}
   \caption{Simulated measurements by STIX and MiSolFA within 1
     min, with $\cos\theta_\text{S}=$ 0.7--0.8 and
     $\cos\theta_\text{M}=$ 0.9--1.0.}
   \label{fig-measured-flux}
\end{figure*}

The most effective way of detecting deviations from a fully isotropic
electron distribution is to look for differences in the shape (not
only in the normalization) of the energy spectrum of the emitted
X-rays.  A sizable difference in the absolute flux reaching the two
detectors is expected.  However, it is very difficult to establish a
standard candle for calibration of absolute fluxes.  On the other
hand, shape differences can be compared against spectra, which can be
safely assumed to have the same shape, for example when the same
source is viewed at about the same angle.  This is why checking for
shape differences is a more robust approach from the experimental
point of view.  The example under consideration, i.e. minimal
variations in anisotropy, represents a challenging case, because it
implies choosing adjacent slices in figure~\ref{fig-true-ratio}, with
minimal shape differences in the energy
spectrum. Figure~\ref{fig-true-flux} shows the corresponding ``true''
energy spectra of the simulated photon flux towards STIX and MiSolFA,
using the same energy binning for both.

After having considered the passage through all materials in front of
the photon detectors, the different energy resolution and background
distributions, one obtains the expected distributions shown in
figure~\ref{fig-expected-flux}.  Next, they are taken as the input for
a pseudo-experiment, in which the expected (real) value in each bin is
taken as the parameter of a Poisson distribution, which is adopted to
generate a random (integer) number of counts in 1 minute of
observation time for both detectors.  The result is shown in
figure~\ref{fig-measured-flux}.

\begin{figure}[t!]
   \centering
   \includegraphics[width=0.9\linewidth]{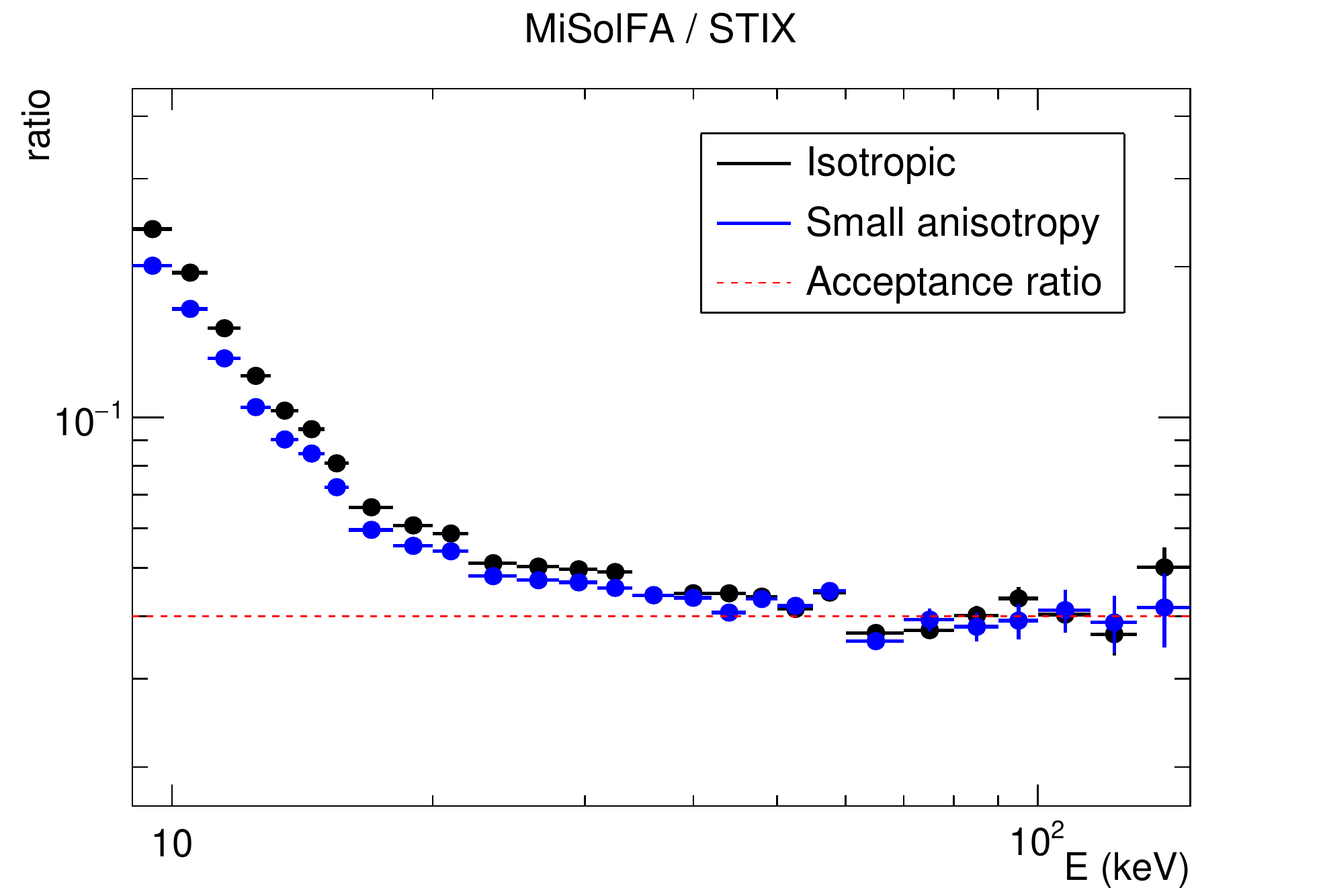}
   \caption{Binwise ratio between the simulated measurements of
     MiSolFA and STIX, for fully isotropic and mildly anisotropic
     parent electron distributions, and a small angular separation
     ($\cos\theta_\text{S}=$ 0.7--0.8 and $\cos\theta_\text{M}=$
     0.9--1.0).}
   \label{fig-ratio}
 \end{figure}

Figures \ref{fig-true-flux} to \ref{fig-measured-flux} show two
distributions for each instrument, one corresponding to the fully
isotropic electron distribution and one obtained with the mild
anisotropy shown in figure \ref{fig-iso-near-iso}.  Error bars reflect
the finite size of the Monte Carlo sample in figures
\ref{fig-true-flux} and \ref{fig-expected-flux}, while
figure~\ref{fig-measured-flux} also includes the uncertainty from the
finite size of the detected sample.  The main goal is to assess how
well these two cases can be distinguished, by comparing the
observations performed with two instruments.  In order to quickly
verify that the two models can indeed be disentangled, we take here
the binwise ratios between the measured distributions by MiSolFA and
STIX and compare these ratios in figure~\ref{fig-ratio}.  These ratios
are clearly different, which implies that the two models can be
distinguished by the simultaneous measurement performed with the two
instruments. Hence, any greater level of flare anisotropy, if present,
will also be measured by both instruments.

Actually, a better approach would be to compare the unfolded
distributions, which are obtained after taking into account all
detector effects and represent our best inference on the ``true''
input of each instrument, and then take the ratio of the unfolded
distributions.  However, the accuracy of the unfolding procedure is
ultimately limited by the statistical uncertainty, which is shown by
the error bars in figure~\ref{fig-ratio}.  Hence the latter provides
sufficient information to verify that the two models produce shape
differences in the measured energy spectra of STIX and MiSolFA.

\section{Expected number of good flares}\label{exp_flares}

Based on the statistical study of three years of RHESSI solar flare
observations performed by \citet{2005A&A...439..737B}, a simulation
was performed, with the purpose of understanding the range of possible
scenarios to be encountered by STIX. Two functional relationships
based on this study have been heuristically obtained from the
simulation.  The first relates the energy threshold $E_\text{thr}$ at
which the non-thermal contribution equals the thermal component to the
flare intensity $I$, taken as the logarithm in base 10 of the GOES
class normalized to X1 (i.e. $I=0$ for X1, $I=-1$ for M1, $I=-2$ for
C1, etc.). A quadratic fit with the function
$E_\text{thr} = p_0 + p_1 I + p_2 I^2$ provides a very good
description of this relationship, with best-fit parameters
$p_0 = 36.3 \pm 0.07$ keV, $p_1 = 3.57 \pm 0.06$ keV, and
$p_2 = -0.301 \pm 0.015$ keV.\ The fit quality is very good, with
$\chi^2 = 0.072$ over 4 degrees of freedom.
However one must be aware that the flare-to-flare variations are so
big that the good fit quality is mostly because of the large spread
among the data.

The second heuristic relationship connects the base-10 logarithm of
the photon rate $R$ in Hz/cm$^{2}$ above $E_\text{thr}$ to the flare
intensity $I$. A parabolic fit with the function $R = q_0 + q_1 I +
q_2 I^2$ provides a very good description of this relationship, with
best-fit parameters 
$q_0 = 3.107 \pm 0.009$,
$q_1 = 1.154 \pm 0.010$, and
$q_2 = 0.059 \pm 0.005$.
The fit quality is very good, with $\chi^2 = 0.008$ over 12 degrees of freedom.
Again, this is mostly a result of the large spread of flare characteristics.

The same relationships can also be exploited for MiSolFA, after
accounting for the instrumental differences mentioned above. 
To put ourselves in the worst case, we considered the detection of a
flare with STIX at perihelion (where it spends only a very short
fraction of its orbit).  With respect to the orbit average, the
closest distance gives a significant intensity magnification, bringing
the ratio between STIX and MiSolFA acceptances to the very
conservative factor of 25, which is used below.
Thus, the rate logarithm $R$ for MiSolFA is taken to be 1.4 units
smaller than STIX, and the expected non-thermal counts for a power law
model can be computed for any given spectral index.
Here we take as an example, a photon spectral index $\gamma=4$ for
each instrument, which is steeper than the photon spectrum considered
above\footnote{Because of the albedo contribution, the photon spectrum is
  not a single power law.  By fitting with this function over
  different energy intervals, one gets a spectral index ranging from
  about 2.6 at lower energies to about 3.2 at higher energies.}, 
to get a conservative estimate of the photon counts. The result
is shown in table~\ref{tab-counts} for MiSolFA.

 \begin{table}
  \caption{Expected non-thermal counts for a photon spectral index of $\gamma=4$
    for MiSolFA. The relative uncertainty is statistical only.}
  \label{tab-counts}
  \centering
  \small 
  \begin{tabular}{c|rrrr|rrrr}
   $E$ bin & \multicolumn{4}{c}{MiSolFA counts/min} &
   \multicolumn{4}{|c}{Relative uncertainty} \\
   (keV)   & M1 & M3 & M5 & X1 & M1 & M3 & M5 & X1 \\
   \hline
   28--31  &  64 &     &     &      & 0.13  \\
   31--34  &  43 &     &     &      & 0.15  \\
   34--38  &  38 & 242 &     &      & 0.16 & 0.06 \\
   38--42  &  25 & 159 & 402 & 1021 & 0.20 & 0.08 & 0.05 & 0.03 \\
   42--46  &  17 & 108 & 274 &  696 & 0.24 & 0.10 & 0.06 & 0.04 \\
   46--50  &  12 &  76 & 192 &  488 & 0.29 & 0.11 & 0.07 & 0.05 \\
   50--55  &  11 &  68 & 171 &  435 & 0.31 & 0.12 & 0.08 & 0.05 \\
   55--60  &   7 &  46 & 118 &  298 & 0.37 & 0.15 & 0.09 & 0.06 \\
   60--70  &   9 &  57 & 145 &  368 & 0.33 & 0.13 & 0.08 & 0.05 \\
   70--80  &   5 &  32 &  82 &  208 & 0.44 & 0.18 & 0.11 & 0.07 \\
   80--90  &   3 &  20 &  50 &  128 & 0.56 & 0.22 & 0.14 & 0.09 \\
   90--100 &   2 &  12 &  31 &   79 & 0.72 & 0.28 & 0.18 & 0.11 \\
  100--115 &   2 &  11 &  29 &   73 & 0.74 & 0.30 & 0.19 & 0.12 \\
  115--130 &   1 &   7 &  18 &   45 & 0.95 & 0.38 & 0.24 & 0.15 \\
   \hline
     sums: & 240 & 840 & 1512 & 3840 \\
  \end{tabular}
 \end{table}

For an M1-class flare, STIX should collect about 6000 non-thermal
counts in one minute at the peak, while MiSolFA in the same time is
expected to see 240 counts.  For an M3-class flare, the expected counts
per minute are 21000 for STIX and 840 for MiSolFA.  For M5 and X1
classes, the expectation is 38k and 96k counts per minute for STIX,
and 1.5k and 3.8k for MiSolFA. As STIX will collect many more events
in each energy bin, the relative uncertainty in the ratio with MiSolFA
is dominated by the counts of the latter instrument.  This
uncertainty (ignoring any systematic effect, which might be discovered
in the future) is also reported in table~\ref{tab-counts}.

All M and X class flares will be suitable for directivity
measurements. For the weaker M-class flares it might be beneficial to
adopt a coarser energy binning, to decrease the statistical
uncertainty in each bin.  However, shape differences like those
expected from the isotropic and mildly anisotropic models considered
above can be measured without rebinning from class M3 above.

The last step is to estimate the expected number of flares of class M1
or higher, at the next solar maximum.  The statistical distribution of
solar flare classes is well described by a power law behaviour, with
spectral index of about 2.1 and percent-level variations in slope
across different solar cycles.
According to the recent review by \citet{2015SpWea..13..286W}, one
would expect to see at least 20 solar flares with class M1 or above
per month during a solar maximum period.  Together with our
conservative estimate of the total overlapping live time of two months
for STIX and MiSolFA, this implies that there should be at least 40
observations suitable for directivity measurement. Observing even one
suitable flare, in which non-isotropic emission is detected and its
dependence on energy is studied, would be a definite step forwards in
our understanding of electron anisotropy in solar flares, and hence
the expected number of good flares is very encouraging.

\section{Summary}\label{summary}

In this work, dual X-ray observations of the same solar flare from two
upcoming instruments, STIX and MiSolFA, at different viewing angles
are considered, in the context of prospective electron directivity
measurements.  A number of instrumental effects have been taken into
account by performing a simulation of the response of STIX and MiSolFA
to the photon fluxes computed in two models, in which the accelerated
electrons are either fully isotropic or close to isotropic, helping us
to determine the capabilities of the instrumentation for electron
directivity measurements. The X-ray albedo component is also taken
into account, as described in \citet{2011A&A...536A..93J}.

Our study focussed on the energy range of the non-thermal component up
to the energy bins in which the background counts are no longer
negligible compared to the photon rate.  Depending on the flare, this
region goes from 20--30 keV to about 100 keV. In order to have enough
counts in MiSolFA, which is the instrument with the smaller
acceptance, flares of class M1 or higher are required, to be able to
distinguish between the two considered models. For such flares, we
find that even for a mildly anisotropic case, and for spacecraft
separations as small as 20--30$^{\circ}$, STIX and MiSolFA will be
able to detect shape differences in the X-ray spectra. Hence, higher
levels of X-ray anisotropy should be easily detectable.
Given the rate of flares as a function of the GOES class, and
conservatively assuming a 8\% net overlapping time for STIX and
MiSolFA, one expects to observe at least 40 flares of M1 class or
above during solar maximum, and it will be possible to perform
quantitative estimates of the X-ray intensity along different
directions, as a function of energy.

Therefore, the result of this study is that there will be at least 40
flares suitable for directivity studies, thanks to the stereoscopic
measurements by STIX and MiSolFA at the next solar maximum. The
combined use of these two instruments will allow for a quantitative
measurement of solar flare electron anisotropy for the first time in
solar flare physics, a vital diagnostic tool for understanding and
constraining fundamental solar flare models of particle acceleration
and transport. 


\begin{acknowledgements}
  The authors wish to thank Gordon Hurford for his help on the
  simulation of the flares distribution for STIX.
  NLSJ and EPK acknowledge support from STFC Consolidated Grant.  The
  NLSJ research leading to these results has received funding from the
  European Community's Seventh Framework Programme (FP7/2007-2013)
  under grant agreement no. 606862 (F-CHROMA).
\end{acknowledgements}


\bibliographystyle{aa}
\bibliography{directivity}

\end{document}